\documentclass[global,twocolumn]{svjour} 
\usepackage[british]{babel}
\usepackage{graphicx}
\usepackage{amsmath}
\usepackage{widetext}
\usepackage{bbm}
\usepackage{amssymb}
\usepackage{varioref}
\usepackage{hyperref}
\usepackage{bibentry}

\usepackage{color}
\usepackage{hyperref}
\usepackage{placeins}

\def\eq{\begin{eqnarray}}
\def\en{\end{eqnarray}}

\newcommand{\ket}[1]{|#1\rangle}
\newcommand{\braket}[2]{\langle{#1}|{#2}\rangle}

\newcommand{\bra}[1]{\langle#1|}

\usepackage[square,numbers,sort&compress]{natbib}
\journalname{Applied Physics B: Lasers and Optics}
\begin{document}
\sloppy
\title{How bosonic is a pair of fermions?}
\author{Malte C.~Tichy\inst{1} \and P.~Alexander Bouvrie\inst{2} \and Klaus M\o{}lmer\inst{1}
}             
\institute{
Department of Physics and Astronomy, University of Aarhus, DK-8000 Aarhus C, Denmark
 \and
Departamento de F\'isica At\'omica, Molecular y Nuclear and Instituto Carlos I de F\'isica Te\'orica y Computacional, Universidad de Granada, E-18071 Granada, Spain
}
\date{\today}
%
\maketitle
\begin{abstract}
Composite particles made of two fermions can be treated as ideal elementary bosons as long as the constituent fermions are sufficiently entangled. In that case, the Pauli principle acting on the parts does not jeopardise the bosonic behaviour of the whole. An indicator for bosonic quality is the \emph{composite boson normalisation ratio} $\chi_{N+1}/\chi_{N}$ of a state of $N$ composites. This quantity is prohibitively complicated  to compute exactly for realistic two-fermion wavefunctions and large composite numbers $N$. Here, we provide an efficient characterisation in terms of the purity $P$ and the largest eigenvalue $\lambda_1$ of the reduced single-fermion state. We find the states that extremise $\chi_N$ for given $P$ and $\lambda_1$, and we provide easily evaluable, saturable upper and lower bounds for the normalisation ratio. Our results strengthen the relationship between the bosonic quality of a composite particle and the entanglement of its constituents. 
\end{abstract}

\section{Introduction}

At all physical scales, bosons and fermions emerge as the two fundamental  species for identical particles, inseparably connected to  their characteristic behaviour: The Pauli principle forbids two fermions to occupy the same state, while bosonic bunching favours such multiple occupation. 

We routinely treat composite particles made of an even number of fermionic constituents as bosons, which seems justified a-posteriori by the success of such  description: From pions composed of two quarks \cite{Collaboration:2010uq} to molecules made of a large number of electrons and nuclei \cite{Zwierlein2003}, bosonic behavior is truly universal. At first sight, however, the Pauli principle that acts on the fermionic parts seems to jeopardise the bosonic behaviour of the whole. Notwithstanding this apparent obstacle, a microscopic theoretical treatment of two-fermion compounds explains the emergence of ideally behaving composite bosons:  A compound of two fermions exhibits bosonic behaviour as long as the constituent fermions are sufficiently entangled \cite{Law2005,Horodecki2009,Tichy2011a}, such that they effectively do not compete for single-fermion states and remain undisturbed by the Pauli principle \cite{Rombouts2002,Sancho2006}. This observation connects our understanding of the almost perfect bosonic behaviour at all scales ranging from sub-nuclear physics \cite{Collaboration:2010uq}  to ultracold molecules \cite{Zwierlein2003} with the tools and concepts of quantum information \cite{Horodecki2009,Gavrilik2012,Gavrilik2013}.

  The \emph{composite boson normalisation ratio} $\chi_{N+1}/\chi_N$  \cite{Law2005,Combescot2011,Combescot2003,Chudzicki2010,Ramanathan2011} of states with  $N+1$ and $N$ \emph{cobosons} (composite bosons)  captures the above argument quantitatively, as we discuss in more detail below. When it is close to unity,  cobosons can be treated as elementary bosons \cite{Law2005}, while deviations are observable in the statistical behaviour of the compounds \cite{avancini,Tichy2012a,Ramanathan2011a,Lee2013,Kaszlikowski2013,Brougham2010,Combescot,Thilagam2013b}. The normalisation factor $\chi_{N}$ depends on the two-fermion wavefunction and answers our above question: ``How bosonic is a pair of  fermions?'' Moreover, the argument can be taken to the realm of Cooper-pairs \cite{Pong2007} and composites made of two elementary bosons, for which a similar analysis is possible \cite{Law2005,TichyBosonsBosons}.  
 
The exact evaluation of $\chi_N$ becomes quickly unfeasible when the number of cobosons $N$ and the number of relevant single-fermion states $S$ are large, which makes approximations desirable. Simple saturable bounds to $\chi_N$ as a function of the purity $P$ of the single-fermion reduced state were derived in Refs.~\cite{Chudzicki2010,Tichy2012CB} and an elegant algebraic approach to prove such bounds was put forward in \cite{Combescot2011}. For very small purities, $P \ll 1/N^2$, the upper and lower bounds converge, which yields an excellent characterisation of the emerging coboson. For moderate values of the purity $P \sim 1/N$, however, a considerable gap between the lower and the upper bound opens up \cite{Tichy2012CB}. In this regime, the $P$-dependent bounds do not characterise the coboson very well, and tighter bounds are desirable.  

Here,  we derive bounds for the normalisation factor $\chi_N$ and for the normalisation ratio $\chi_{N+1}/\chi_N$ for two-fermion cobosons which depend on the purity $P$ \emph{and} on the largest eigenvalue $\lambda_1$ of the single-fermion density matrix $\hat \rho_{(a)}$, introduced below.  The bounds can be evaluated efficiently for very large composite numbers $N$, and we show that they permit a significantly more precise characterisation of two-fermion cobosons than bounds in $P$ alone \cite{Chudzicki2010,Tichy2012CB}. 

We introduce the physics of cobosons and motivate the importance of the normalisation ratio in Section \ref{microdecr}. Our main result, a set of saturable bounds for the normalisation ratio, is derived in Section \ref{boundsformulation}. Examples and a discussion of the bounds are given in Section \ref{discex}. An outlook on possible future developments that take into account further characteristics of the wavefunction is given in Section \ref{conclout}. Technical details regarding the derivation of the bounds are given in the Appendices \ref{App1} and \ref{app2}.

\section{Algebraic description of cobosons} \label{microdecr}

\subsection{Coboson normalisation factor}
 We consider two distinguishable\footnote{In our context, a state of two indistinguishable fermions can always be mapped formally  onto distinguishable fermions \cite{Tichy2012CB}. Our subsequent discussion therefore applies  to distinguishable and indistinguishable fermions in a similar fashion. For composites made of two bosons, however, we  expect differences between distinguishable and indistinguishable bosons due to multiply populated single-boson states \cite{TichyBosonsBosons}. }  fermions of species $a$ and $b$ prepared in a collective wavefunction of the form  \eq \ket{\Psi}= \sum_{j,k=1}^{\infty} \omega_{j,k} \ket{A_j, B_k} , \en 
where we assume that the two-fermion state can be expanded on a discrete set of single-fermion states, which is fulfilled for bound states and also incorporates possible spin-coupling. The bases $\{ \ket{A_j} \}$, $\{ \ket{B_k} \}$ can be chosen at will, and it is convenient to use the \emph{Schmidt decomposition} of $\ket{\Psi}$ \cite{Bernstein2009}, i.e.~to choose two particular single-particle bases $\ket{a_j}$ and $\ket{b_j}$ with 
\eq 
\ket{\Psi}= \sum_{j=1}^S \sqrt{\lambda_j} \ket{a_j , b_j} ,  \label{SchmidtDecomp}\\
 \lambda_1 \ge \lambda_2 \ge \dots  \ge 0, ~~ \sum_{j=1}^S \lambda_j =1,   \label{eq:ddefin}
\en
where the ordering of the $S$ Schmidt coefficients $\lambda_j$ is imposed for convenience such that $\lambda_1$ be the largest coefficient in the distribution $\vec \Lambda=(\lambda_1, \dots , \lambda_S)$, and $S$ is not necessarily finite. 
The $\lambda_j$ coincide with the eigenvalues of either reduced single-fermion  density matrix, 
\eq 
\hat \rho_{(a)}=\sum_{j=1}^S \lambda_j \ket{a_j}\bra{a_j}, ~~~ \hat \rho_{(b)}=\sum_{j=1}^S \lambda_j \ket{b_j}\bra{b_j} . 
\en 
 We treat a pair of fermions in the state $\ket{\Psi}$ as a coboson, for which we can define an approximate creation operator in second quantization \cite{Law2005} 
\eq 
\hat c^\dagger=\sum_{j=1}^S \sqrt{\lambda_j} ~ \hat a^\dagger_j \hat b^\dagger_j =: \sum_{j=1}^S \sqrt{\lambda_j} ~ \hat d^\dagger_j  \label{eq:cdefin} , 
\en
where $\hat a^\dagger_j$ ($\hat b^\dagger_j$) creates a fermion in the Schmidt-mode $\ket{a_j}$ $(\ket{b_j})$. 
 The operator $\hat d_j^\dagger$ creates a \emph{bi-fermion} in a product state, i.e.~a pair of two fermions in their respective mode $j$. While such creation and annihilation operators commute, 
\eq
\left[\hat d_j , \hat d_k \right] &=& \left[\hat d_j^\dagger , \hat d_k^\dagger \right] =0  , 
\en 
bi-fermions also obey the Pauli principle, such that
\eq 
\left( \hat d_j^\dagger \right)^2 &=& \left( \hat d_j \right)^2 =0 . \label{pauli}
\en
 As a consequence, the operators $\hat c, \hat c^\dagger$ do not fulfil the ideal bosonic commutation relation, but obey \cite{Law2005} 
\eq [\hat c, \hat c^\dagger] = 1 -  \sum_{j=1}^S \lambda_j  ( \hat a_j^\dagger \hat a_j +  \hat b_j^\dagger \hat b_j  ) .  \label{commu1} \en 

An $N$-coboson state is  obtained by the $N$-fold application of the creation operator (\ref{eq:cdefin}) on the vacuum \cite{Law2005}, 
\eq 
\ket{N} =\frac{  \left( \hat c^\dagger \right)^N }{ \sqrt{\chi^{\vec \Lambda}_N~ N!} }\ket{0}  , \label{NCompositeState}
\en
where $\chi_N^{\vec \Lambda} \le 1$ is the coboson \emph{normalisation factor} \cite{Combescot2003,Law2005,Combescot2008a}, which ensures that $\ket{N}$ is normalised to unity. Inserting the definition of the coboson creation operator (\ref{eq:cdefin}) into (\ref{NCompositeState}), we find 
\eq 
\ket{N} =\frac{1}{\sqrt{\chi_N^{\vec \Lambda} N!}}  \sum_{j_1 \neq j_2 \dots \neq j_N}^{1 \le j_m \le S} \prod_{k=1}^N  \sqrt{\lambda_{j_k}} \hat d^\dagger_{j_k} ,  \label{superposrep}
\en
where terms with repeated indices $j_{m}=j_k$ do not contribute, due to the Pauli principle ensured by Eq.~(\ref{pauli}). In other words, the $N$-coboson state is a superposition of $N$ bi-fermions that are distributed among the bi-fermion Schmidt modes. Each distribution of the bi-fermions in the modes is weighted by $N!$ coherently superposed amplitudes.

\subsection{Algebraic properties of the normalisation factor}
By evaluating the norm of the $N$-coboson state in Eq.~(\ref{superposrep}), one obtains a closed expression for the coboson normalisation factor $\chi_N^{\vec \Lambda}$ as the elementary symmetric polynomial of degree $N$ in the Schmidt coefficients $\vec \Lambda$ \cite{MacDonaldSymmetric}:
\eq 
\chi_N^{\vec \Lambda}&=&  \Omega \{ \underbrace{1 \dots 1}_{N} \}  ,  \label{chifdef} \\
\Omega \{ x_1 \dots x_N  \} &=& N! \sum_{1 \le p_1 < \dots < p_N \le S} \prod_{k=1}^N \lambda_{p_k}^{x_k} , \label{chiBdefini} 
\en
where the latter can be expressed recursively
\eq 
\Omega \{ x, \underbrace{1 \dots 1}_{K} \}  \label{Recursion}
&=& M(x) { \Omega}\{ \underbrace{  1 \dots 1 }_{K} \}  -  K{ \Omega}\{ x+1, \underbrace{1\dots 1 }_{K-1} \} , ~~~~
\en
with the help of the power-sums of order 1 to $N$, 
\eq 
M(k) = \sum_{j=1}^S \lambda_j^k, ~~~ M(2)\equiv P, ~~~ M(1)=1 .\en
 Alternatively, the Newton-Girard identities \cite{Ramanathan2011,MacDonaldSymmetric} can be used, 
\eq 
 \chi^{\vec \Lambda}_N&=&(N-1)! \sum_{m=1}^{N} \frac{(-1)^{1+m}\chi^{\vec \Lambda}_{N-m}}{(N-m)!} M(m) , \label{NewtonGirard}
\en
which are more suitable in practice than Eqs.~(\ref{chiBdefini},\ref{Recursion}). 

The computation of $\chi_N$ becomes significantly simpler when all Schmidt coefficients in a distribution $\vec \Lambda$ are identical. In this case, all summands in Eq.~(\ref{chiBdefini}) are equal, and counting the number of terms gives
\eq 
\chi_N^{\vec \Lambda}  &=& \lambda^{N} \frac{S!}{(S-N)!} \label{singlelam} ~ \text{ for } \vec \Lambda= (\underbrace{ \lambda  \dots \lambda }_{S}) , \label{samelambdaD}
\en
which can be combined with \cite{TichyBosonsBosons}
\eq
\chi_N^{(\lambda_1 \dots \lambda_S ) } & = & \sum_{M=0}^N \chi_M^{( \lambda_1  \dots \lambda_L )} ~ \chi_{N-M}^{ ( \lambda_{L+1}  \dots \lambda_S )}  { N \choose M } , \label{binorec}  
\en
to quickly yield $\chi_N^{\vec \Lambda}$ for distributions $\vec \Lambda$ with large Schmidt coefficient multiplicities. 

\subsection{Bosonic behaviour in relation to the normalisation ratio}
The \emph{normalisation ratio} $\chi_{N+1}^{\vec \Lambda}/\chi_N^{\vec \Lambda}$ \cite{Law2005} determines the bosonic quality of a state of $N$  cobosons. For an intuitive picture, consider one summand in Eq.~(\ref{superposrep}), in which the $N$ bi-fermions occupy the modes $j_1, \dots, j_N$. In order to add an $N+1$st coboson to the state $\ket{N}$, we need to accommodate it among the $S-N$ unoccupied Schmidt modes. The probability that the added bi-fermion successfully ends up in an unoccupied Schmidt mode is then the sum of the coefficients associated to these unoccupied modes, $\sum_{m \notin \{ j_1, \dots, j_N  \} }  \lambda_m$. This argument can be repeated for each configuration $j_1, \dots, j_N$, and the success probability to add an $N+1$st coboson to an $N$-coboson state becomes 
\eq
\frac 1 {\chi_N^{\vec \Lambda}} \sum_{j_1 \neq j_2 \dots \neq j_N}^{1 \le j_m \le S} \prod_{k=1}^N  \lambda_{j_k}  \left[ \sum_{m \notin \{ j_1, \dots, j_N  \} }  \lambda_m \right] \nonumber \\
 =
\frac 1 {\chi_N^{\vec \Lambda}}  \sum_{j_1 \neq j_2 \dots \neq j_N \neq j_{N+1}}^{1 \le j_m \le S} \prod_{k=1}^{N+1}  \lambda_{j_k}  = \frac {\chi_{N+1}^{\vec \Lambda}} {\chi_N^{\vec \Lambda}}   ,
\en
which is reflected by the sub-normalisation of the state obtained upon application of the creation operator $\hat c^\dagger$ on the $N$-coboson state  \cite{Chudzicki2010}
\eq 
\hat c^\dagger \ket{N} = \sqrt{\frac{\chi_{N+1}^{\vec \Lambda}}{\chi_N^{\vec \Lambda}}} \sqrt{N+1} \ket{N+1} .  \label{addp}
\en

On the other hand, the annihilation of a coboson in an $N$-coboson state yields a state that contains a component orthogonal to the $(N-1)$-coboson state \cite{Law2005},  
\eq 
\hat c \ket{N} =\sqrt{ \frac{\chi_N^{\vec \Lambda}}{\chi_{N-1}^{\vec \Lambda}}} \sqrt{N} \ket{N-1} + \ket{\epsilon_N} ,  \label{subaddp}
\en
with 
\eq 
\braket{\epsilon_N}{\epsilon_N}=1-N \frac{\chi_N^{\vec \Lambda}}{\chi_{N-1}^{\vec \Lambda}} + (N-1) \frac{\chi_{N+1}^{\vec \Lambda}}{\chi_{N}^{\vec \Lambda}} .
\en
Combining the relations (\ref{addp}) and (\ref{subaddp}), one finds the expectation value of the commutator   (\ref{commu1}) on an $N$-coboson state \cite{Law2005,Ramanathan2011,Combescot2011}, 
\eq
\bra N  \left[ \hat c, \hat c^\dagger \right] \ket N&=& 1 -  2\sum_{j=1}^S \lambda_j   \bra N \hat n_j  \ket N \nonumber \\
&=&2 \frac{\chi^{\vec \Lambda}_{N+1}}{\chi^{\vec \Lambda}_N} -1 , \label{commutatorexpl}
 \en
 where $\hat n_j=\hat d^\dagger_j \hat d_j$ counts the number of bi-fermions in mode $j$. For an ideal boson, Eq.~(\ref{commutatorexpl}) will equate to unity. Since all observable bosonic behavior is borne by the bosonic commutation relations, values of $\chi_{N+1}^{\vec \Lambda}/\chi_N^{\vec \Lambda}$ close to unity witness a statistical behaviour that is close to the ideal bosonic one, while deviations from unity come with observable consequences that are induced  by the statistics of the constituent fermions \cite{Combescot,Ramanathan2011a,Tichy2012a,Thilagam2013,Lee2013}. 

\section{Bounds on the normalisation factor and ratio} \label{boundsformulation}
Given a wavefunction $\ket{\Psi}$ of two distinguishable fermions, one can, in principle, diagonalise one reduced single-fermion density matrix $\hat \rho_{(a/b)}$ to obtain the distribution $\vec \Lambda$, and compute $\chi_N$ with the help of the previous formulae, Eqs.~(\ref{chifdef},\ref{chiBdefini},\ref{NewtonGirard},\ref{samelambdaD},\ref{binorec}). 

In practice, however, even if the full distribution $\vec \Lambda$ or all relevant power-sums $M(2) \dots M(N)$ are actually known, the evaluation of the normalisation factor $\chi_N$ is unfeasible for very large numbers of cobosons: Using Eq.~(\ref{NewtonGirard}), for example, the computation of $\chi_N$ requires the knowledge of all $\chi_M$ with $M<N$.  Already for  a harmonically trapped condensate of hydrogen atoms, the exact approach turns out to be unfeasible \cite{Chudzicki2010}.  

A characterisation of $\chi_N$ in terms of few, well-accessible quantities, such as the largest eigenvalue $\lambda_1$ and the purity $P$ of the reduced single-fermion density matrix is therefore essential in practice.  The largest eigenvalue can be approximated via power iteration \cite{Bernstein2009}, while the purity is basis-independent and fulfils $P=\text{Tr}[\hat \rho_{(a)}^2]=\text{Tr}[\hat \rho_{(b)}^2]$. Full diagonalisation of $\hat \rho_{(a/b)}$ is not necessary for either quantity, while both bear clear physical meaning as quantifier of entanglement: The \emph{Schmidt number} \cite{Grobe:1994kl,Horodecki2009} is defined as $K=1/P$, the \emph{geometric measure of entanglement} \cite{PRA68Godbart}  fulfils  $E_{G}=1-\lambda_1$. 

Upper and lower bounds to the normalisation factor $\chi_N^{\vec \Lambda}$ and  to the normalisation ratio $\chi_{N+1}^{\vec \Lambda}/\chi_{N}^{\vec \Lambda}$ in terms of  $P$ and $\lambda_1$ are therefore highly desirable, not only to permit the efficient evaluation of $\chi_N$ in practice, but also to provide a better physical understanding of the connection between quantum entanglement and bosonic behavior. 

Bounds as a function of the single-fermion purity $P\equiv M(2)=\text{Tr}(\hat \rho_{(a/b)})$ were put forward previously \cite{Chudzicki2010,Ramanathan2011,Tichy2012CB,Combescot2011}. In the regime $P \ll 1/N^2$, the bounds are  efficient and tightly confine the possible values of $\chi_N$ and $\chi_{N+1}/\chi_N$ \cite{Tichy2012CB}. For moderate values of $P \sim 1/N$, however, the upper and lower bounds differ considerably, i.e.~the compounds are not well-characterised by $P$ alone, and higher-order power-sums $M(m \ge 3)$ become important in the expansion in Eq.~(\ref{NewtonGirard}). 

Here, we formulate bounds that depend on $P$ \emph{and} on the largest Schmidt coefficient $\lambda_1$. Existing bounds \cite{Chudzicki2010,Combescot2011,Tichy2012CB} emerge naturally as extremal cases in the limit of the minimal and maximal value of $\lambda_1$ for a given $P$. The extremal distributions of Schmidt coefficients that emerge below coincide with the ones derived in Ref.~\cite{TichyBosonsBosons} for two-boson composites. The alternating sign in Eq.~(\ref{NewtonGirard}), however, has no analogy for two-boson compounds, such that the approach of Ref.~\cite{TichyBosonsBosons} needs to be adapted to fit the present case.  

\subsection{Lower bound in $P$ and $\lambda_1$}
We assume that we are given a distribution $\vec \Lambda$ with largest Schmidt coefficient $\lambda_1$ and purity $P$. The distribution $\vec \Lambda_{\text{min}}(P, \lambda_1)$  that minimises $\chi_N$ under these constraints is derived in Appendix \ref{App1}.

The resulting \emph{minimising distribution} $\vec \Lambda_{\text{min}}(P, \lambda_1)$ \cite{TichyBosonsBosons} contains $S$ non-vanishing Schmidt coefficients, with ${\lambda_1 \ge \lambda_2 = \lambda_3 = \dots = \lambda_{S-1} \ge \lambda_S}$, and 
\eq 
S&=&1 + \left \lceil \frac{(1-\lambda_1)^2}{P-\lambda_1^2} \right \rceil  \label{Sdefintext}. 
\en 
The normalisation in Eq.~(\ref{eq:ddefin}) and the fixed purity $P$ imply for the Schmidt coefficients $\lambda_j$ 
\eq 
 \lambda_1 + (S-2) \lambda_2  + \lambda_S &=& 1 , \nonumber \\ 
\lambda_1^2 + (S-2) \lambda_2^2 +  \lambda_S^2 &=&  P  \label{q2}  .
\en
With
\eq 
R= \sqrt{(S - 2) ( \lambda_1 (2 - S \lambda_1) + (S - 1) P - 1)} ,
\en
the relevant solution to Eq.~(\ref{q2}) is \cite{TichyBosonsBosons}
\eq 
\lambda_{2, \dots, S-1}&=& \frac{1-\lambda_1}{S-1} + \frac{R}{(S-2)(S-1)} ,  \nonumber  \\
\lambda_S&= & \frac{1-\lambda_1-R}{S-1}  , \label{LowerBoundLambdaSintext}
\en
where $\lambda_1 \ge \lambda_2 \ge \lambda_S$ is fulfilled by construction. 

\begin{widetext}
Given such distribution of three distinct Schmidt coefficients $\lambda_1, \lambda_2, \lambda_S$ with multiplicities $1, S-2, 1$, respectively, we can compute $\chi_N^{\vec \Lambda_{\text{min}}(P, \lambda_1)}$ using Eqs.~(\ref{samelambdaD},\ref{binorec}): 
\eq 
\chi_N^{\vec \Lambda_{\text{min}}(P,\lambda_1)}  =  \lambda_2^{N-2}\left[ (N - S) \lambda_2  ((N - S+1) \lambda_2 -   N (\lambda_1 + \lambda_S) ) + (N-1 ) N \lambda_1 \lambda_S \right] \frac{(S-2)!}{(S-N)! } ,   \label{eq:minboundSN} \en
where we used $1/k! = 0 $ for $k<0$. Given $\lambda_1$ and $P$, this expression can be readily evaluated, even for large values of $N$. 

Consistent with the Pauli-principle, it is impossible to populate $S$ Schmidt modes with $N>S$ bi-fermions, which is ensured by the factor $1/(S-N)!$ in Eq.~(\ref{eq:minboundSN}). In general, Eq.~(\ref{Sdefintext}) imposes
\eq 
2 + \left \lceil \frac{(1-\lambda_1)^2}{(P-\lambda_1^2)} \right\rceil \le N \Rightarrow    \chi_N^{\vec \Lambda_{\text{min}}(\lambda_1, P) }  = 0 . \label{mindS}
\en

The normalisation \emph{ratio} ${\chi_{N+1}^{\vec \Lambda_{\text{min}}(P,\lambda_1)} }/ {\chi_{N}^{\vec \Lambda_{\text{min}}(P,\lambda_1)}}$  is a monotonically increasing function of $S$. We can therefore obtain a simpler, however slightly weaker, lower bound for Eq.~(\ref{eq:minboundSN}) by setting $S=1+(1-\lambda_1)^2/(P-\lambda_1^2)$, i.e. we omit the ceiling-function in Eq.~(\ref{Sdefintext}): 
\eq 
\chi_N^{\vec \Lambda_{\text{min}}(\lambda_1, P) } \ge 
 \frac{  \Gamma\left[  \frac{ (1- \lambda_1)^2}{P - \lambda_1^2} \right] }{ \Gamma\left[2 - N +  \frac{ (1- \lambda_1)^2}{P - \lambda_1^2} \right] } \left(1 + \left(N-2 \right) \lambda_1- P \left(N-1\right) \right) \left( \frac{P - \lambda_1^2}{1 - \lambda_1} \right)^{N-2} , \label{eq:smoothlower}
\en
which is only applicable for $1 + \left\lceil \frac{(1-\lambda_1)^2}{(P-\lambda_1^2)} \right\rceil \ge N $ [see Eq.~(\ref{mindS})]. 
For values of $P$ and $\lambda_1$ for which $(1-\lambda_1)^2/(P-\lambda_1^2) $ is integer, the smooth lower bound Eq.~(\ref{eq:smoothlower}) exactly coincides with the exact expression, Eq.~(\ref{eq:minboundSN}). 
 \end{widetext}

\subsection{Upper bound in $P$ and $\lambda_1$} 
In strict analogy to the last section, we construct the distribution $\vec \Lambda_{\text{max}}(\lambda_1, P)$  that \emph{maximises} the normalisation constant $\chi_N$ for fixed $\lambda_1$ and $P$ in Appendix \ref{app2}. 

In $\vec \Lambda_{\text{max}}(\lambda_1, P)$ \cite{TichyBosonsBosons}, the multiplicity of $\lambda_1$  is chosen as large as possible, i.e. $\lambda_1$ is repeated $L-1$ times, with $L=\lceil P/\lambda_1^2 \rceil$. The $L$th coefficient is then maximised, while the remaining $S-L$ coefficients fulfil
${\lambda_1=\lambda_2=\dots = \lambda_{L-1} \ge  \lambda_L \ge \lambda_{L+1} = \dots = \lambda_S  }$. To ensure normalisation [Eq.~(\ref{eq:ddefin})] and satisfy  $M(2)=P$, we have
\eq 
(L-1) \lambda_1 + \lambda_L + (S-L) \lambda_S &=&1 , \nonumber \\
(L-1) \lambda_1^2 + \lambda_L^2 + (S-L) \lambda_S^2 &=&P  .
\en
With 
\eq 
R^\prime = \nonumber
\sqrt{( S-L) ( P (S+1- L)-1 + ( L-1) \lambda_1 (2 - \lambda_1 S))} ,
\en
we find the relevant solution for $\lambda_L$ and $\lambda_S$ \cite{TichyBosonsBosons}, 
\eq 
\lambda_L&=&  \frac{ 1-(L-1)\lambda_1+R^\prime}{S+1-L} , \nonumber \\ 
\lambda_{S}&=& \frac{1-(L-1)\lambda_1}{S+1-L} -\frac{R^\prime }{(S-L)(S+1-L)} , \label{MaxLambdaSDef}
\en
where, in order to ensure $\lambda_S, \lambda_L \ge 0$, $S$ needs to fulfil 
\eq 
S > \frac{(L-1)P+1-2(L-1)\lambda_1}{P-(L-1)\lambda_1^2}.
\en

Using Eqs.~(\ref{binorec},\ref{samelambdaD}), the normalisation factor for the maximising distribution becomes 

\begin{widetext}

\eq \nonumber
\chi_N^{\vec \Lambda_{\text{max}}(P, \lambda_1)} 
& \stackrel{\text{Eq.}(\ref{binorec})}{=} & \sum_{K=0}^N \sum_{M=0}^{N-K} \chi^{(\lambda_1, \dots, \lambda_1)}_{M} \chi^{(\lambda_L)}_{K}  \chi^{(\lambda_S, \dots, \lambda_S)}_{N-M-K}  {N \choose M, K} \\
& \stackrel{\text{Eq.}(\ref{samelambdaD}) }{=} & \sum_{K=0}^1 \sum_{M=0}^{N-K}  \frac{(L-1)!}{(L-1-M)!}  \frac{(S-L)!}{(S-L-(N-K-M))!} \lambda_1^{M} \lambda_L^K  \lambda_S^{N-M-K}  {N \choose M, K}  ,
\en
where ${ X \choose Y, Z}=\frac{X!}{Y! Z! (X-Y-Z)!}$ is the multinomial coefficient. 
Since this expression is an increasing function of $S$, we maximise it in the limit $S \rightarrow \infty$. Defining $\lambda_\Sigma$ as the sum of all infinitesimal coefficients $\lambda_S$ in that limit, we find 
\eq
\lambda_\Sigma=(1-(L-1)\lambda_1-\sqrt{\lambda_1^2(1-L)+P}), ~~~~~~~ \lim_{S\to\infty} \chi^{(\lambda_S, \dots, \lambda_S)}_{N-M-K}=   \lambda_\Sigma^{N-M-K} , 
\en
which gives
\eq \nonumber
\chi_N^{\vec \Lambda_{\text{max}}(P, \lambda_1)}  
\label{ChiMaxUsingLimit}
& = & \sum_{K=0}^1 \sum_{M=0}^{N-K}  \frac{(L-1)!}{(L-1-M)!}   \lambda_1^{M} \lambda_L^K  \lambda_\Sigma^{N-M-K} {N \choose M, K} \\
& = &(-\lambda_1)^{L-1}\lambda_\Sigma^{N-L} \left[N \lambda_L ~\mathcal{U}\left(1-L,1-L+N,-\frac{\lambda_\Sigma}{\lambda_1}\right) + \lambda_\Sigma ~\mathcal{U}\left(1-L,2-L+N,-\frac{\lambda_\Sigma}{\lambda_1}\right)\right], \label{ULdef}
\en
where $\mathcal{U}(a,b,z)$ is Tricomi's confluent hypergeometric function \cite{Olver:2010:NHMF}, which allows fast numerical evaluation in practice. Using $\lambda_1 \ge \lambda_L$, we find a simpler upper bound to the above expression: 
\eq
\chi_N^{\vec \Lambda_{\text{max}}(P, \lambda_1)}   \le  \sum_{M=0}^{\min(N,\lfloor \tilde L \rfloor +1 )}  \frac{ \Gamma\left( \tilde L+1 \right)}{\Gamma\left( \tilde L-M+1 \right) }   \lambda_1^{M}   \lambda_\Sigma^{N-M} {N \choose M} , \label{upperboundsmooth}
\en
where $\tilde L=P/\lambda_1^2$ (note the omitted ceiling-function), and $\lambda_\Sigma$ is evaluated for $L=\tilde L$. The last expression coincides with Eq.~(\ref{ULdef}) when $P/\lambda_1^2$ is integer, since in that case $L=\tilde L$ and $\lambda_L=\lambda_1$. 

\end{widetext}

\begin{figure}[ht]
\includegraphics[width=\linewidth]{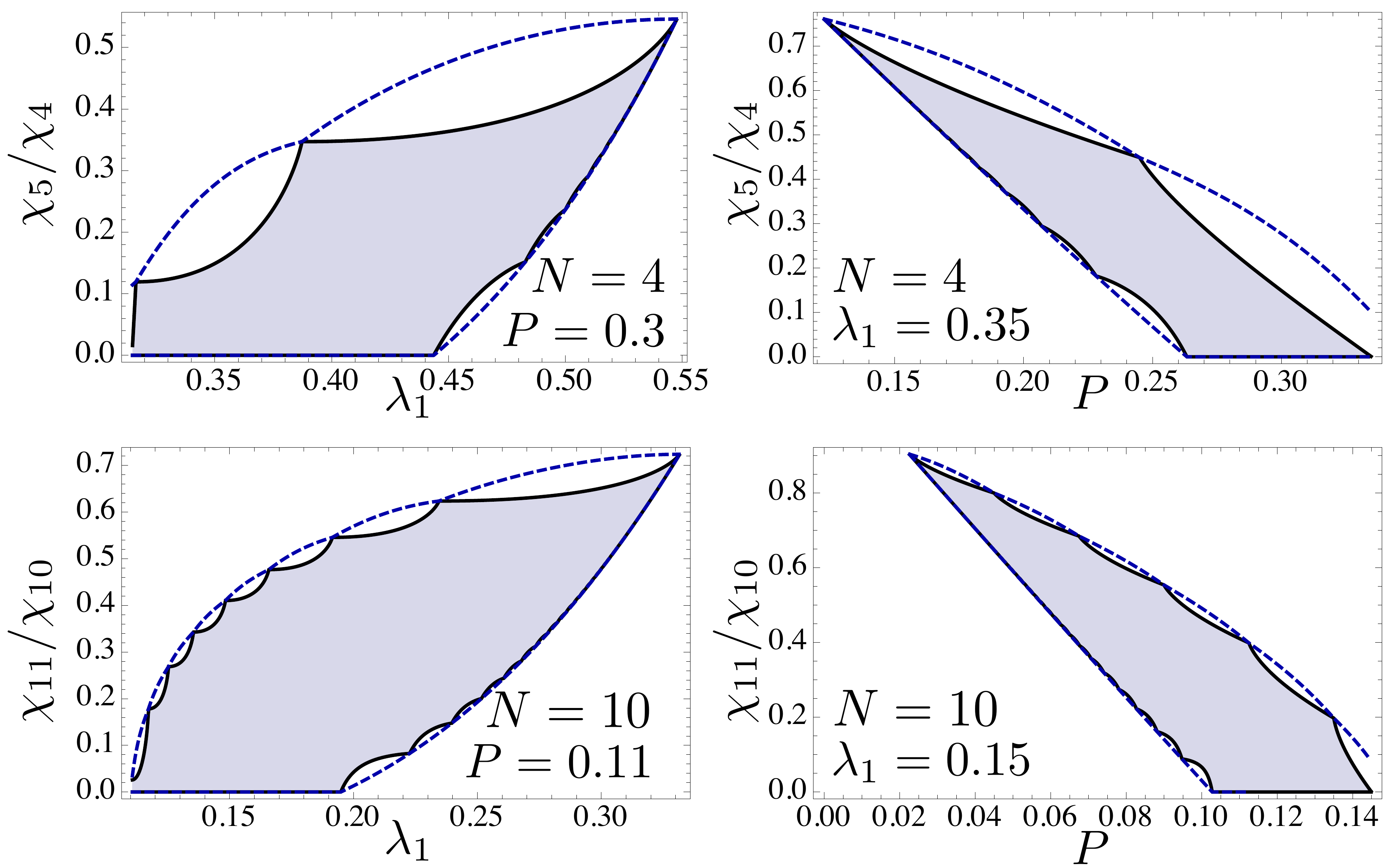}
\caption{Exact values of the normalisation ratio $\chi_{N+1}/\chi_N$ for the minimising and maximising distribution [black solid lines, computed using Eqs.~(\ref{eq:minboundSN},\ref{ULdef})] and smooth upper and lower bounds [blue dashed lines, Eqs.~(\ref{eq:smoothlower},\ref{upperboundsmooth})]. Upper row: $N=4$, lower row: $N=10$. Left column: Fixed purity $P$, the normalisation ratio is shown as a function of the largest Schmidt coefficient $\lambda_1$. Right column: Fixed $\lambda_1$, the normalisation ratio is shown as a function of $P$. The normalisation ratio of any distribution $\vec \Lambda$ with $P$ and $\lambda_1$ is restricted to the shaded range delimited by the black solid line. }
\label{BoundsSmooth.pdf}
\end{figure}

We compare the tight saturable bounds, Eqs.~(\ref{eq:minboundSN}) and (\ref{ULdef}), with their respective smooth approximations, Eqs.~(\ref{eq:smoothlower}) and (\ref{upperboundsmooth}), in Fig.~\ref{BoundsSmooth.pdf}. 

\subsection{Bounds in $P$}
The parameters $\lambda_1$ and $P$ cannot be chosen independently, since, by construction \cite{TichyBosonsBosons}, 
\eq 
P \le \lambda_{1,\text{min}}(P) & \le&  \lambda_1  \le \lambda_{1,\text{max}}(P) =  \sqrt{P} ,  \label{constraintsl}
\en
where 
\eq 
\label{lambdaminP}
\lambda_{1,\text{min}}(P) &=& \frac{1}{\left\lceil\frac 1 P \right\rceil } \left( \sqrt{\frac{P \left\lceil\frac 1 P \right\rceil -1}{{\left\lceil\frac 1 P \right\rceil -1}}}+1 \right) .
\en
We obtain $P$-dependent and $\lambda_1$-independent upper (lower) bounds to $\chi_N$ and $\chi_{N+1}/\chi_N$ by fixing $P$ and setting the largest Schmidt coefficient to its extremal value, $\lambda_{1,\text{max(min)}}(P)$.

\subsubsection{Upper bound in $P$}
We maximise the normalisation factor and ratio by choosing $\lambda_{1}=\lambda_{1,\text{max}}(P)=\sqrt{P}$.  The minimising distribution $\vec{\Lambda}_{\text{min}}(P,\lambda_1)$ and the maximising distribution $\vec{\Lambda}_{\text{max}}(P,\lambda_1)$  then both converge to the \emph{peaked} distribution \cite{Tichy2012CB}, $\mathbf{\Lambda}_\text{peak}(P)$, given by the limit $S\rightarrow \infty$ of 
\eq
\label{PeakDistributionP}
\lambda_{1,\text{peak}} &=& \frac{1 + \sqrt{(S-1)(S P -1)}}{S},  \nonumber \\
\lambda_{j \in \{ 2\dots S\},\text{peak}} &=& \frac{1-\lambda_{1,\text{peak}}}{S-1} .
\en
Via Eqs.~(\ref{samelambdaD},\ref{binorec}), we recover the $P$-dependent upper bound \cite{Tichy2012CB}
\eq
 \chi_N^{\vec{\Lambda}_\text{peak}(P)} = (1-\sqrt{P})^{N-1} \left[ 1+(N-1) \sqrt{P} \right].~~ \label{eq:upperboundP} 
\en

\subsubsection{Lower bound in $P$}
The normalisation factor and ratio are minimised for fixed $P$ by choosing ${\lambda_{1}=\lambda_{1,\text{min}}(P)}$, as given by Eq.~\eqref{lambdaminP}. In this case,  both distributions $\vec{\Lambda}_{\text{min}}(P,\lambda_1)$ and $\vec{\Lambda}_{\text{max}}(P,\lambda_1)$ become the \emph{uniform}  distribution \cite{Tichy2012CB}, $\vec{\Lambda}_\text{uni}(P)$, with $S=L=\left\lceil \frac{1}{P} \right\rceil$ non-vanishing Schmidt coefficients given by 
\eq
\lambda_{j \in \{ 1\dots L-1\},\text{uni}} &=& \lambda_{1,\text{min}}(P) , \nonumber \\ 
\lambda_{L,\text{uni}} & =& 1-\lambda_{1,\text{min}}(P) (L-1) .  
\label{UniformDistributionP}
\en
Using Eqs.~(\ref{samelambdaD},\ref{binorec}), we recover the lower bound \cite{Tichy2012CB} 
\eq
 \chi_N^{\vec{\Lambda}_\text{uni}(P)} =   \frac{\lambda_{1,\text{uni}}^{N-1}  (L-1)! }{(L-N)! } \left[
N-L \lambda_{1,\text{uni}}(N-1)
   \right] . \label{eq:lowerboundP}
\en

\subsection{Bounds in $\lambda_1$}
The constraints on $\lambda_1$ and $P$ in Eq.~(\ref{constraintsl}) can be re-formulated as constraints on $P$:
\eq
 \lambda_1^2 = P_{\text{min}}(\lambda_1) &\le&  P \le P_{\text{max}}(\lambda_1) \le \lambda_1  
\label{PExtremeValues}
\en
where
\eq
 P_{\text{max}}(\lambda_1) &=&\lambda_1^2 \left\lfloor \frac 1 {\lambda_1} \right\rfloor + \left( 1- \lambda_1 \left\lfloor \frac 1 {\lambda_1} \right\rfloor  \right)^2 \label{PAndLambdaMaxMin} .
\en
We obtain $\lambda_1$-dependent and $P$-independent upper (lower) bounds to the normalisation ratio and factor by choosing $P_{\text{min(max)}}(\lambda_1)$.

\subsubsection{Upper bound in $\lambda_1$}
For $P= P_{\text{min}}(\lambda_1)$, the distributions $\vec{\Lambda}_{\text{min/max}}(P,\lambda_1)$ become a \emph{peaked} distribution, $\vec{\Lambda}_{\text{peak}}(P_{\text{min}}(\lambda_1))$, with the first Schmidt coefficient $\lambda_1$ and ($S-1$) coefficients of magnitude $(1-\lambda_1)/(S-1)$. In the limit $S\to\infty$ the normalisation factor reads
\eq
\chi_N^{\vec{\Lambda}_\text{peak}(P_{\text{min}}(\lambda_1))} = (1-\lambda_1)^{N-1} (1+(N-1) \lambda_1) .~~  \label{onlyl1a} 
\en
Since $\lambda_1 \le \sqrt{P} $, this upper bound is always larger (i.e.~weaker) than the upper bound in $P$ given by Eq.~(\ref{eq:upperboundP}): 
\eq 
\chi_N^{\vec{\Lambda}_\text{peak}(P_{\text{min}}(\lambda_1))} \ge \chi_N^{\vec{\Lambda}_\text{peak}(P)} , \label{upperweaker}
\en
for any pair  $(P, \lambda_1)$ fulfilling  Eq.~(\ref{constraintsl}).

\begin{figure*}
\includegraphics[width=1\textwidth]{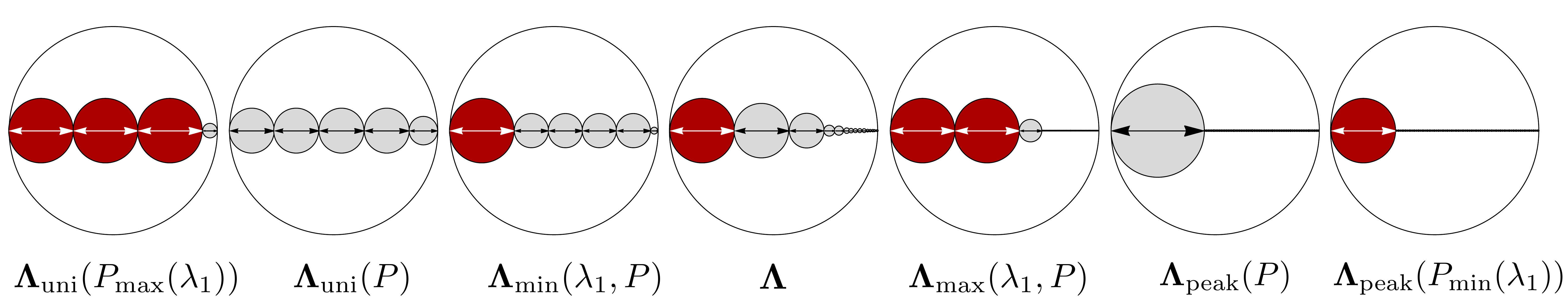} 
\caption{Hierarchy of minimising and maximising distributions. The circle diameters correspond to the magnitude of a Schmidt coefficient $\lambda_j$, and the fraction of filled area in each large circle is the purity $P$ of the respective distribution. A distribution $\vec \Lambda$, with $\lambda_1=0.3$ and $P=0.2$ (center) leads to a normalisation factor $\chi_N$ that is bound from below and from above by the $\chi_N$ evaluated for the distributions on the left and on the right, respectively. The order of the distributions reflects the hierarchy of Eq.~(\ref{BoundsSummaries}). All circles that correspond to $\lambda_1$ are filled with dark red and marked with white arrows. The resulting normalisation ratios $\chi_{N+1}/\chi_N$ obey the same hierarchy, as illustrated by  the intersections of the vertical lines $(i)$ with the three minimising and the three maximising limits in Fig.~\ref{BoundsFunctionPL1}. The normalisation ratio of $\vec \Lambda$ then lies on $(i)$ within the shaded area.}
\label{FigureDistributions.pdf}
\end{figure*}

\subsubsection{Lower bound in $\lambda_1$}
We find a lower bound in $\lambda_1$ by setting $P = P_{\text{max}}(\lambda_1)$, as given by Eq.~\eqref{PAndLambdaMaxMin}.  The resulting distribution contains the largest possible multiplicity of $\lambda_1$, i.e. it contains $L-1=S-1=\left\lfloor\frac{1}{\lambda_1}\right\rfloor$ coefficients of magnitude $\lambda_1$ and one of magnitude $\left(1-\left\lfloor\frac{1}{\lambda_1}\right\rfloor \lambda_1 \right)$. The resulting normalisation factor fulfils
\eq
\chi_N^{\vec{\Lambda}_\text{uni}(P_{\text{max}}(\lambda_1))} 
= \frac{\lambda_1^{N-1}  (L-1)! }{\left(L-N \right)!  }\left[N- \lambda_1 L (N-1)  \right] .  \label{onlyl1b}
\en
In analogy to Eq.~(\ref{upperweaker}), this lower bound in $\lambda_1$ is always smaller (i.e.~weaker) than the corresponding bound in $P$: 
\eq 
\chi_N^{\vec{\Lambda}_\text{uni}(P_{\text{max}}(\lambda_1))} \le \chi_N^{\vec{\Lambda}_\text{uni}(P)} ,
\en
due to $P \le P_{\text{max}}$. 

\section{Summaries of the bounds and discussion} \label{discex}
Examples for all pertinent distributions are shown in Fig.~\ref{FigureDistributions.pdf}: A randomly chosen distribution $\vec \Lambda$ (middle panel) with specified $\lambda_1$ and $P$ leads to a certain normalisation factor $\chi_N^{\vec \Lambda}$, which is bound from below by the distributions on the left and from above by those from the right, successively. Summarising the attained values for the normalisation factor given in Eqs.~(\ref{eq:minboundSN},\ref{ULdef},\ref{eq:upperboundP},\ref{eq:lowerboundP},\ref{onlyl1a},\ref{onlyl1b}), we obtain our main result, 
\eq 
\chi_N^{\vec \Lambda_{\text{uni}}(P_{\text{max}}(\lambda_{1}))}  \le 
\chi_N^{\vec \Lambda_{\text{uni}}(P)} \le  
\chi_N^{\vec \Lambda_{\text{min}}(\lambda_1, P) }  \nonumber \hspace{2cm} \\
\hspace{2cm} \le\chi_N^{\vec \Lambda} \le \hspace{3.4cm} \label{BoundsSummaries} \\
\hspace{2cm} \chi_N^{\vec \Lambda_{\text{max}}(\lambda_1, P) }\le 
\chi_N^{\vec \Lambda_{\text{peak}}(P)} \le 
\chi_N^{\vec \Lambda_{\text{peak}}(P_{\text{min}}(\lambda_1) )  } \nonumber .
\en
This hierarchy of consecutively tighter bounds is immediately inherited by the normalisation ratio $\chi_{N+1}/\chi_N$ in full analogy, which quantitatively answers our initial question, ``How bosonic is a pair of fermions?'', in terms of $P$ and $\lambda_1$. 

In order to obtain a physical understanding of these bounds, a combinatorial approach is instructive: The normalisation factor $\chi_N$ can be interpreted as the probability that a collection of $N$ objects that are each given a property $j$ with probability $\lambda_j$ does not contain any set of two or more objects with the same property \cite{Tichy2012CB} (for $S=365$ and $\lambda_j=1/365$, we recover the ``birthday problem'' \cite{Munford1977}). 
In our physical context, no two or more bi-fermions are allowed to occupy the same Schmidt mode. The Pauli principle, enforced by Eq.~(\ref{pauli}), implies that the emerging $N$-coboson state in Eq.~(\ref{superposrep}) does not contain any such terms describing multiple occupation. The lack of these terms then needs to be accounted for by the normalisation factor $\chi_N$. 
 
\subsection{Entanglement and bosonic behavior}
 \begin{figure}[ht]
\includegraphics[width=.5\textwidth]{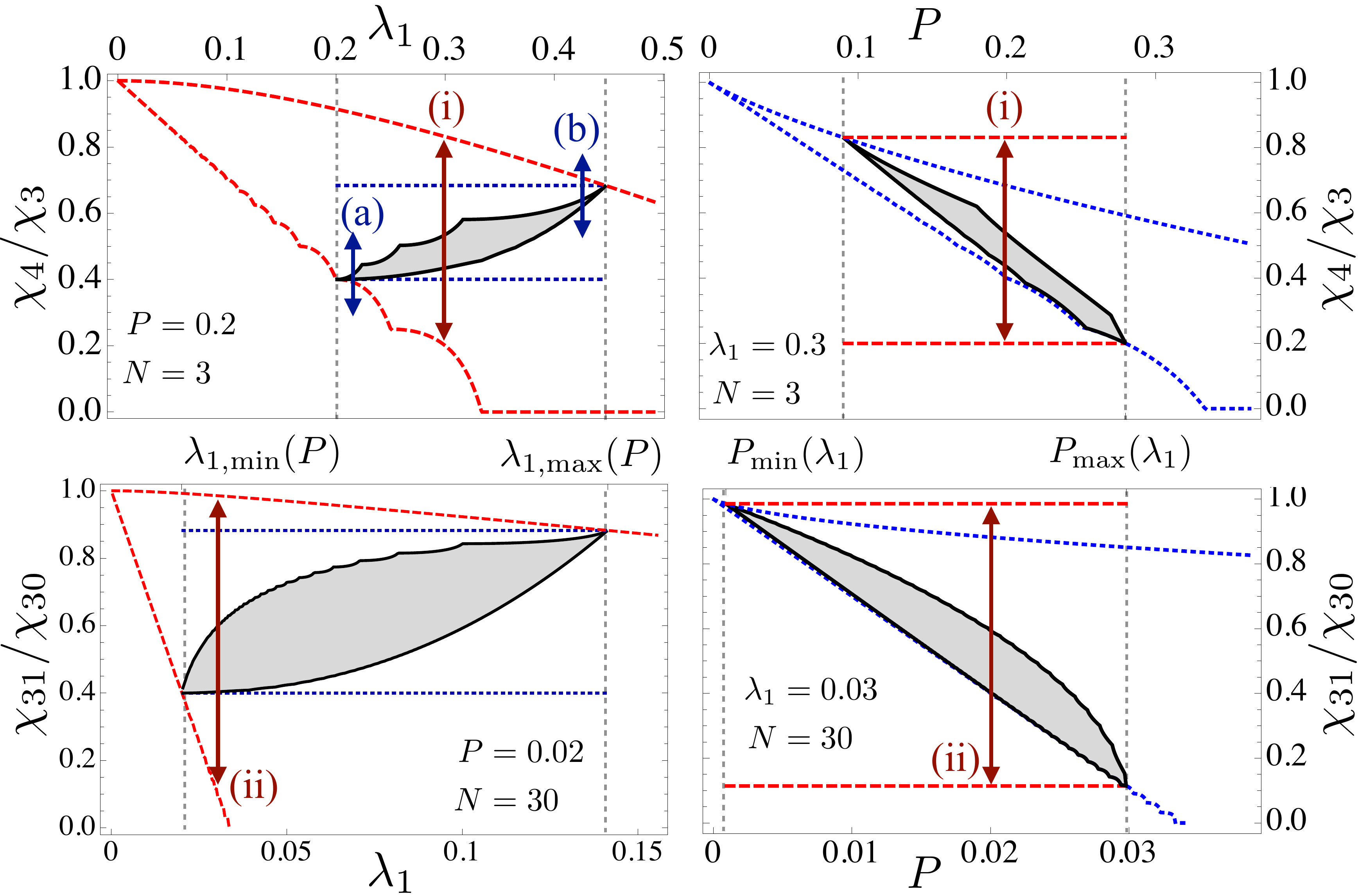}
\caption{Upper and lower bounds to the normalisation ratio $\chi_{N+1}/\chi_N$ as a function of $\lambda_1$ (left panels) and $P$ (right panels), for $N=3$ (top row) and $N=30$ (bottom row). Red dashed lines correspond to bounds in $\lambda_1$ alone, Eqs.~(\ref{onlyl1a},\ref{onlyl1b}); blue dotted lines show the bounds in $P$ alone, Eq.~(\ref{eq:upperboundP},\ref{eq:lowerboundP}). The combined bounds, Eqs.~(\ref{eq:minboundSN},\ref{ULdef}), are shown as solid black lines, the shaded area is the range allowed for general distributions $\vec \Lambda$ with given $\lambda_1$ and $P$. The bounds in $P$ are always superior to those in $\lambda_1$. By setting $P$ (left panel) or $\lambda_1$ (right panel), the possible values of $\lambda_1$ and $P$, respectively, are constrained by Eqs.~(\ref{constraintsl}) and (\ref{PExtremeValues}). The solid vertical lines in the upper panels indicate those values for which the maximising and minimising distributions are depicted in Fig.~\ref{FigureDistributions.pdf} [solid red lines (i), $\lambda_1=0.3, P=0.2$] and Fig.~\ref{SpecialDistr.pdf} [solid dark blue lines, $P=0.2$, (a) $\lambda_1=0.215$, (b) $\lambda_1=0.42$]. The vertical lines (ii) in the lower panel indicate corresponding values of $P$ and $\lambda_1$. }
\label{BoundsFunctionPL1}
\end{figure}
 
Combinatorially speaking, the purity $P$ represents the probability that two randomly chosen objects possess the same property (it is therefore also called the collision entropy). Here, it reflects the probability that the wavefunction vanishes upon two bi-fermions competing for the same Schmidt mode.  Therefore, the $P$-dependent bounds on $\chi_N$ decrease monotonically with increasing $P$ (blue dotted lines in the right panels of Fig.~\ref{BoundsFunctionPL1}). Larger entanglement, characterised by a smaller purity $P$, is therefore tantamount to a more bosonic composite \cite{Law2005,Chudzicki2010,Tichy2012CB}.

Similarly,  the $\lambda_1$-dependent bounds decrease with increasing $\lambda_1$ (red dashed lines in the left panels of Fig.~\ref{BoundsFunctionPL1}). Consistently, an increase of $\lambda_1$ also leads to weaker geometric entanglement, $E_G=1-\lambda_1$. This connection underlines, again, the relationship between quantum entanglement and the bosonic behavior of composites. 

The knowledge of $\lambda_1$ alone leaves a finite range for possible values of $P$ [see Eq.~(\ref{PExtremeValues})]: The remaining, unknown Schmidt coefficients $\lambda_2\dots \lambda_S$ may be many and small, or few and large (compare the distribution $\vec \Lambda_{\text{uni}}(P_{\text{max}}(\lambda_1))$ to $\vec \Lambda_{\text{peak}}(P_{\text{min}}(\lambda_1))$ in Fig.~\ref{FigureDistributions.pdf}). Indeed, the main sources of deviation from bosonic behavior are binary ``collisions'' of bi-fermions,  which is directly quantified by $P$. Therefore, bounds in $\lambda_1$ are always weaker than bounds in $P$; in the formalism of quantum information, the purity $P$ is more decisive than the overlap with the closest separable state, $\lambda_1$. 

\begin{figure}[ht]
\includegraphics[width=.5\textwidth]{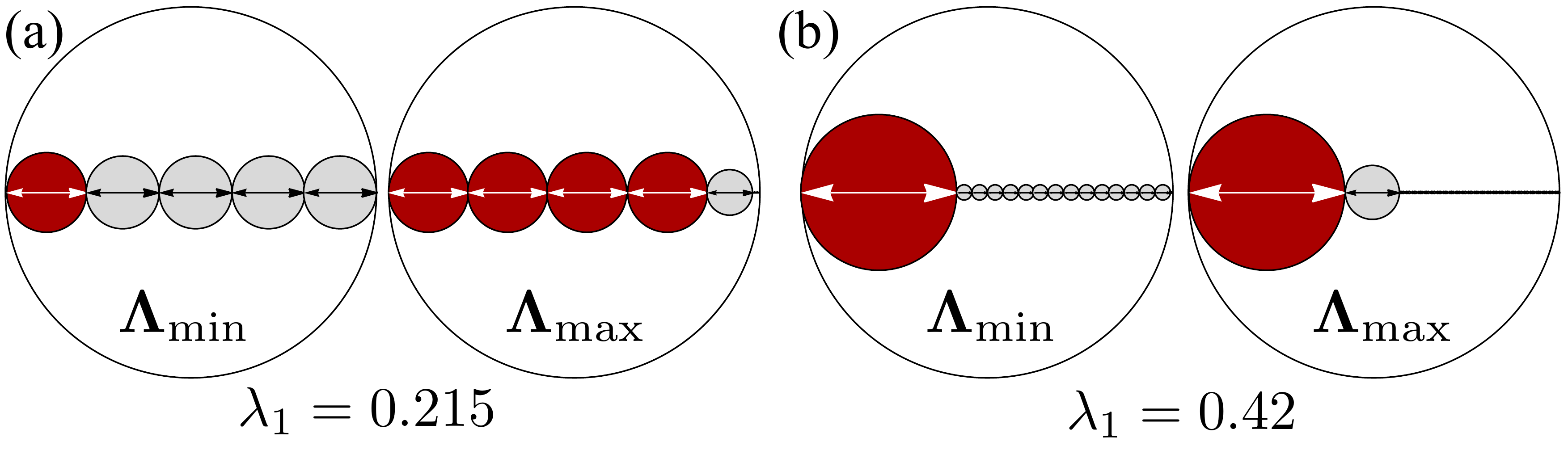}
\caption{Minimising and maximising distributions $\vec \Lambda_{\textrm{min/max}}$,
 for close-to-extremal values of $\lambda_1$ and fixed $P=0.2$. (a) $\lambda_1=0.215 \gtrapprox \lambda_{1,\text{min}}(P)$. (b) $\lambda_1=0.42 \lessapprox  \lambda_{1,\text{max}}(P)$.  The emerging bounds correspond to the black solid lines in Fig.~\ref{BoundsFunctionPL1} at the intersections with arrows (a) and (b), respectively. For $\lambda_1 \rightarrow \lambda_{1,\textrm{max(min)}}(P)$, the distributions converge to the peaked (uniform) distribution (compare to the corresponding sketches in Fig.~\ref{FigureDistributions.pdf}). }
\label{SpecialDistr.pdf}
\end{figure}

The knowledge of both, $\lambda_1$ and $P$, yields a considerable enhancement over bounds in $P$ alone (black solid lines in Fig.~\ref{BoundsFunctionPL1}). In particular,  the range of possible $\chi_N$ becomes narrower for extremal values of $P$ or $\lambda_1$, for which the minimising and maximising distributions resemble each other, as in Fig.~\ref{SpecialDistr.pdf}. In this case, $\lambda_1$ and $P$ strongly constrain the remaining Schmidt coefficients. 

In view of the clear dependence of $\chi_N$ on $P$ and $\lambda_1$, it is remarkable that the \emph{combined} bound in $P$ and $\lambda_1$ features an \emph{increase} of the bosonic quality $\chi_N$ and $\chi_{N+1}/\chi_N$ with $\lambda_1$ (Fig.~\ref{BoundsFunctionPL1}).  This increase, however, is due to the fixed purity $P$: By increasing the largest Schmidt coefficient $\lambda_1$, all other Schmidt coefficients need to decrease in order to keep $P$ constant, which naturally increases the total accessible number of Schmidt modes, and, consequently, $\chi_N$. More formally speaking, $\chi_N$ actually \emph{increases} with $M(3)$, as can be inferred  from Eqs.~(\ref{Recursion},\ref{NewtonGirard}) \cite{Ramanathan2011}.

\begin{figure}[ht]
\includegraphics[width=.5\textwidth]{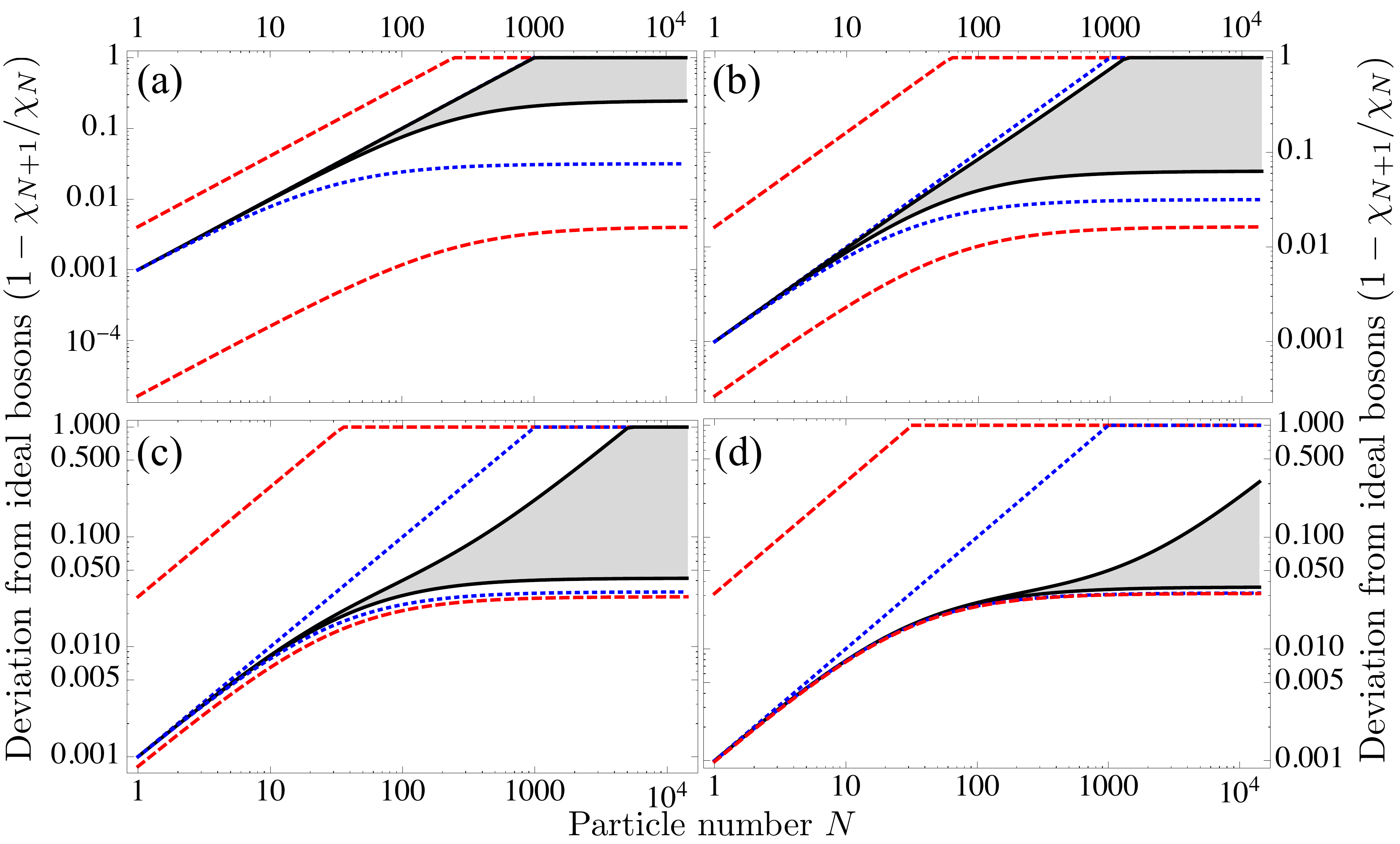}
\caption{Upper and lower bounds to $(1-\chi_{N+1}/\chi_N)$, i.e. to the deviation from ideal bosonic behaviour, as a function of  $N$. The color-code is the same as in Fig.~\ref{BoundsFunctionPL1}. In all panels, $P=0.001$, i.e. the bounds in $P$ alone (blue dotted) do not change.  We choose different values of $\lambda_1$: \hspace{4cm}
$\text{(a) }  {\lambda_1=0.9 \lambda_{1,\text{min}}(P) +  0.1 \lambda_{1,\text{max}}(P) \approx 0.0041}$.
$\text{(b) } {\lambda_1=0.5  \lambda_{1,\text{min}}(P) + 0.5 \lambda_{1,\text{max}}(P) \approx 0.0163}$.
$\text{(c) } {\lambda_1=0.1 \lambda_{1,\text{min}}(P) + 0.9 \lambda_{1,\text{max}}(P) \approx 0.0286}$.
$\text{(d) }  {\lambda_1=0.01 \lambda_{1,\text{min}}(P) + 0.99 \lambda_{1,\text{max}}(P) \approx 0.0313}$.
}
\label{FunctionOfN.pdf}
\end{figure}

\subsection{Limit of large coboson numbers $N$}
In Fig.~\ref{FunctionOfN.pdf}, we show the deviation from the ideal value $\chi_{N+1}/\chi_N=1$ as a function of the number of cobosons $N$. While the upper and lower bounds in $P$ converge for small values of $N < 1/\sqrt{P}$, bounds in $\lambda_1$ do not: For small particle numbers, the coboson behaviour is essentially defined by the binary collision probability, i.e. by the purity $P$. The magnitude of the largest Schmidt coefficient $\lambda_1$ is secondary. For large particle numbers $N > 1/\sqrt{P}$, the knowledge of $\lambda_1$ then fixes the possible range of $M(3)$, which constrains the accessible values of the normalisation ratio. Again, very large or very small values of $\lambda_1$ lead to a tighter confinement of the range of possible $\chi_{N+1}/\chi_N$ than intermediate values of $\lambda_1$, as can be seen by comparing the panels in Fig.~\ref{FunctionOfN.pdf}. In general, $\lambda_1$ and $P$ determine to a wide extent up to which number of cobosons $N$ a condensate of two-fermion composites still behaves bosonically \cite{Combescot2008,Rombouts2002}. 

In comparison to the bounds on the normalisation factor for cobosons made of two elementary bosons \cite{TichyBosonsBosons}, the role of the $\lambda_1$-dependent bounds is exchanged: for two-fermion cobosons,  $\chi_N$ is maximised (minimised) by choosing the smallest (largest) possible purity for a given $\lambda_1$; for two-boson cobosons, the normalisation factor instead \emph{increases} with the  purity. As a consequence, the clear hierarchy of bounds expressed by Eq.~(\ref{BoundsSummaries}) is absent for two-boson cobosons \cite{TichyBosonsBosons}.  This dependence is due to the possibility for multiple occupation of Schmidt modes by bosonic constituents, forbidden by the Pauli principle for fermionic constituents. Furthermore,  when the number of cobosons $N$ is large, $N> 1/\lambda_1$,  the behaviour of two-boson bosons is very well defined by $\lambda_1$ alone, and the multiple occupation of the most prominent Schmidt mode dominates the picture, a process without analogy in the present two-fermion case.


\section{Conclusions and outlook} \label{conclout}
Starting with the general  description  of a two-fermion composite in Eqs.~(\ref{SchmidtDecomp},\ref{eq:cdefin}), we confined the quantitative indicator $\chi_{N+1}/\chi_N$ for the bosonic behaviour of the resulting coboson. For a fixed purity $P$, the immediate difference between the state that minimises and the state that maximises $\chi_N$ is the magnitude of the largest Schmidt coefficient, which is of the order of $P$ for the minimal, uniform distribution, and $\sqrt P$ for the maximal, peaked distribution \cite{Tichy2012CB}. Therefore, the additional constraint on $\lambda_1$ can considerably enhance $P$-dependent bounds \cite{Tichy2012CB,Chudzicki2010,Combescot2011}. 

Our bounds strengthen the relation between quantum entanglement and the bosonic quality of bi-fermion pairs, first established in Ref.~\cite{Law2005}: Not only is the purity $P$ a quantitative indicator for bosonic behavior \cite{Law2005,Chudzicki2010,Ramanathan2011,Tichy2012CB}, but so is the geometric measure of entanglement  \cite{PRA68Godbart}, which can be expressed here  as a function of $\lambda_1$. 

Depending on the application, the single-fermion purity $P$, the largest eigenvalue $\lambda_1$ of the single-fermion density matrix $\hat \rho_{(a/b)}$, or both may be known. We can formulate a clear hierarchy: Knowledge of $P$ is more valuable than the knowledge of $\lambda_1$ alone, whereas the combination can greatly enhance the bounds, depending on the value of the involved parameters. The effect of compositeness are observable in any physical observable that is affected by the commutation relation (\ref{commutatorexpl}), such as, e.g., bosonic signatures in multiparticle interference \cite{Tichy2012a}. 

Our method can be extended to  formulate even stronger bounds that depend on the purity $P$ and on the $m$ largest Schmidt coefficients $\lambda_1 \ge \lambda_2 \ge \dots \ge \lambda_m$: In close analogy to the procedure in \cite{Tichy2012CB} (see Appendix \ref{App1} and \ref{app2}), minimising and maximising distributions can be constructed, and the resulting normalisation factors can be computed. The increased accuracy will, however,  come at the expense of an increased computational cost, since a larger number of distinct Schmidt coefficients (up to $m+2$ when we fix the $m$ largest coefficients and the purity $P$) also leads to a larger number of sums when Eq.~(\ref{binorec}) is applied. 

Another desideratum is the extension of the present bounds to multi-fermion systems in order to characterise, e.g., $\alpha$-particles in extreme environments \cite{Funaki2009,Zinner2008}. The absence of the Schmidt decomposition, Eq.~(\ref{SchmidtDecomp}), for multipartite states \cite{Horodecki2009} makes this task, however, rather challenging. In particular, a simple combinatorial interpretation of the normalisation constant seems to be excluded for such composites.

\section*{Acknowledgments} 
The authors would like to thank Florian Mintert, \L{}ukasz Rudnicki, Alagu Thilagam and Nikolaj Th. Zinner for stimulating discussions, and Christian K. Andersen, Durga Dasari, Jake Gulliksen, Pinja Haikka, David Petrosyan and Andrew C. J. Wade for valuable feedback on the manuscript. M.C.T. gratefully acknowledges support by the Alexander von Humboldt-Foundation through a Feodor Lynen Fellowship. K.M. gratefully acknowledges support by the Villum Foundation. P.A.B. gratefully acknowledges support by the Progama de Movilidad Internacional CEI BioTic en el marco PAP-Erasmus. 

\appendix

\section{Appendix: Minimising distribution} 
\label{App1}

For completeness, we reproduce the proofs from the Appendix of Ref.~\cite{Tichy2012CB}, adapting the argument to our situation in which not only the purity $P$ is fixed, but also the largest Schmidt coefficient $\lambda_1$. 

\subsection{Uniforming operation}
Following an analysis of the birthday-problem with non-uniform birthday probabilities \cite{Munford1977}, we define a \emph{uniforming}  operation $\Gamma^{u}$ on the distribution $\vec \Lambda$ that can modify three selected $\lambda_j$ with indices ${2 \le j_1 < j_2 < j_3 \le S}$ (i.e. the operation never acts on the first Schmidt coefficient $\lambda_1$, since its value is fixed, by assumption). We will show that this operation always decreases $\chi_N$, and specify the distribution $\vec \Lambda_{\text{min}}(P, \lambda_1)$ that remains invariant under the application of $\Gamma^u$. This distribution thus minimises $\chi_N^{\vec \Lambda}$ under the constraints $(P,\lambda_1)$. 
 
The operation $\Gamma^u$ modifies three coefficients in a distribution, 
\eq 
\Gamma^u: ( \lambda_{j_1}, \lambda_{j_2},\lambda_{j_3} ) \rightarrow ( \lambda^u_{j_1}, \lambda^u_{j_2},\lambda^u_{j_3} ) ,
\en
such that it leaves 
\eq
K_1&=&\lambda_{j_1} + \lambda_{j_2} + \lambda_{j_3} , \nonumber \\ 
K_2&=&\lambda_{j_1}^2 + \lambda_{j_2}^2 + \lambda_{j_3}^2 ,\label{K2d}
 \en
 invariant, and, consequently, also $\sum_{j} \lambda_j=1$ and $\sum_j \lambda_j^2=P$. The third power-sum, $M(3)=\sum_j \lambda_j^3$, on the other hand, is changed by $\Gamma^u$. Specifically, 
 \eq 
\lambda_{j_1}^u = \lambda_{j_2}^u &=& \frac{1}{6} \left(2 K_1 + \sqrt{6 K_2-2 K_1^2} \right) \nonumber  ,\\
\lambda_{j_3}^u&=&\frac{1}{3} \left(K_1 - \sqrt{6 K_2-2 K_1^2} \right) .  \label{ope1} 
\en
In the case $K_1^2 <2 K_2$, in order to avoid $\lambda_{j_3}^u<0$, we need to set
\eq 
\lambda_{j_1/j_2}^u&=&\frac{1}{2} \left( K_1 \pm \sqrt{2 K_2-K_1^2}\right) , \label{ope2}\\
\lambda_{j_3}^u&=&0 \nonumber . 
\en

\subsubsection{The product  $\lambda_{j_1} \lambda_{j_2}\lambda_{j_3}$ decreases under $\Gamma^u$} \label{proofproduct}
It holds 
\eq 
\lambda_{j_1}^u \lambda_{j_2}^u \lambda_{j_3}^u 
 \le \lambda_{j_1} \lambda_{j_2} \lambda_{j_3} . \label{lambdaprohier1}
\en

{\it Proof:} We write the left- and right-hand side of (\ref{lambdaprohier1}) in terms of $K_1, K_2$ and $\lambda_{j_1}$
\eq 
\lambda_{j_1}^u \lambda_{j_2}^u \lambda_{j_3}^u = \hspace{6cm} \nonumber \\
=  \left\{ \begin{tabular}{ll}
 $\frac{1}{108} \left(K_1-\sqrt{6 K_2-2 K_1^2}\right) $  &  \\ 
$\times \left(2 K_1+\sqrt{6 K_2-2 K_1^2}\right)^2$ &for $ K_1^2 > 2 K_2$   \vspace{0.24cm}  \\ 
0 & for $ K_1^2 \le 2 K_2$  \end{tabular} \right. \\
 \lambda_{j_1} \lambda_{j_2} \lambda_{j_3}= \frac{1}{2} \lambda_{j_1} \left(2 \lambda_{j_1}^2-2 \lambda_{j_1} K_1+K_1^2-K_2 \right)  \nonumber 
\en

Given $K_1$ and $K_2$, the original $\lambda_{j_{2}/j_{3}}$ become functions of $\lambda_{j_1}$,
\eq
\lambda_{{j_2}/{j_3}} = \frac 1 2 \left( K_1 - \lambda_{j_1} \pm \sqrt{2 \lambda_{j_1} K_1 - K_1^2 - 3 \lambda_{j_1}^2 + 2 K_2} \right) \nonumber  . \en
The requirement $\lambda_{j_1} \ge \lambda_{{j_2}} \ge \lambda_{{j_3}} \ge0$ imposes 
\eq
\frac {  K_1}{3}  + \frac{ \sqrt{3 K_2 - K_1^2}} {3\sqrt{2}}  &\le& \lambda_{j_1} \le \frac { K_1}{3} +\frac{  \sqrt{6 K_2 -2 K_1^2} }{3} .
\en
The values of $\lambda_1$ constrained to this interval then fulfil Eq.~(\ref{lambdaprohier1}).  \qed

\subsubsection{$\chi_N^{\vec \Lambda}$ decreases upon application of $\Gamma^u$} \label{proofdecreases}
Upon application of $\Gamma^u$, the normalisation constant $\chi_N$ and the normalisation ratio $\chi_{N+1}/\chi_N$ can only decrease: 
\eq 
\chi_{N}^{\Gamma^{u}(\vec \Lambda) } & \le&  \chi_{N}^{\vec \Lambda}  \label{operationinequality} , \\
\frac{\chi_{N+1}^{\Gamma^{u}(\vec \Lambda) } }{\chi_{N}^{\Gamma^{u}(\vec \Lambda) }  } & \le& 
\frac{ \chi_{N+1}^{\vec \Lambda} }{ \chi_{N}^{\vec \Lambda}} \label{operationinequality2} .
\en

{\it Proof:} To ease notation in the following, we exemplarily choose  $j_1=2, j_2=3, j_3=4$ and set 
\eq 
\tilde \chi_N&=&\chi_N^{(\lambda_1, \lambda_5, \dots, \lambda_S)},  
\en
which allows us to write $\chi_N$ as 
\eq \chi_N &=&  \lambda_2  \lambda_3  \lambda_4 \cdot  \tilde \chi_{N-3} +(\lambda_2 \lambda_3 +\lambda_4 \lambda_3+\lambda_2 \lambda_4)  \tilde \chi_{N-2}  \nonumber \\
&& 
 +(\lambda_2+\lambda_3+\lambda_4) \tilde \chi_{N-1}+ \tilde \chi_{N} . \label{rewritechn} \en
The terms \eq 
\lambda_2 \lambda_3 +\lambda_4 \lambda_3 +\lambda_2 \lambda_4 &=& \frac 1 2 \left( K_1^2 - K_2 \right) , \\ 
\lambda_2+\lambda_3+\lambda_4&=& K_1 ,   \en 
and $\tilde \chi_{k \in \{ N-3, \dots ,  N \} }$ remain invariant under the application of $\Gamma^{u}$, whereas  the product $ \lambda_2  \lambda_3  \lambda_4$ decreases, due to Eq.~(\ref{lambdaprohier1}). Consequently, also $\chi_N^{\vec \Lambda}$ decreases upon the application of $\Gamma^u$.  

Using $\chi_{N+1} \le  \chi_{N}$, one can easily show in analogy to Ref.~\cite{Tichy2012CB} that the inequality (\ref{operationinequality}) is inherited by the normalisation ratio (\ref{operationinequality2}). 
 \qed

\subsection{Properties of the minimising distribution}
\label{SubSubSecMinimisingDistribution}
The distribution $\vec \Lambda_{\text{min}}(P,\lambda_1)$ that minimises $\chi_N$ for fixed $\lambda_1$ and $P$ should remain invariant under the application of $\Gamma^u$, for all choices of $2 \le j_1, j_2, j_3 \le S$. By the definition of $\Gamma^u$, we see that any three coefficients with $\lambda_{j_1} > \lambda_{j_2} = \lambda_{j_3}$ never constitute a fixed point of $\Gamma^u$. Therefore, the invariant distribution is of the form
\eq
 \lambda_1 \ge \lambda_2=\dots = \lambda_{S-1} \ge  \lambda_S . 
 \en
It coincides with the distribution found in Ref.~\cite{TichyBosonsBosons}.

\section{Appendix: Maximising distribution} \label{app2}

\subsection{Peaking operation}
With $K_1$ and $K_2$ defined as in Eq.~(\ref{K2d}) above, we define the \emph{peaking} operation $\Gamma^p$ as follows \cite{Tichy2012CB}:
 For $K_1+\sqrt{6 K_2-2K_1^2} \le 3 \lambda_1$, we set
\eq 
\lambda_{j_1}^{p}&=&\frac{1}{3} \left(K_1 + \sqrt{6 K_2-2 K_1^2} \right) , \nonumber \\
\lambda_{j_2}^{p}=\lambda_{j_3}^{p} &=& \frac{1}{6} \left(2 K_1 - \sqrt{6 K_2-2 K_1^2} \right)   .~~~~~
\en
If $K_1+\sqrt{6 K_2-2K_1^2} > 3 \lambda_1$, the above definition leads to $\lambda_{j_1}^{p}> \lambda_1$, which we excluded by assumption. In this case, we  define alternatively
\eq 
\lambda_{j_1}^{p}&= & \lambda_1 ,  \\
\lambda_{j_2/j_3}^{p} &=& \frac{ K_1 - \lambda_1 \pm \sqrt{2 (K_2 -\lambda_1^2) -(K_1-\lambda_1)^2 
} }{{2}}  ,   \nonumber 
\en
for which $\lambda_1 = \lambda_{j_1}^p \ge \lambda_{j_2}^p \ge \lambda_{j_3}^p \ge 0 $. 
In full analogy to the discussion in Sections \ref{proofproduct}, \ref{proofdecreases}, one  shows that 
\eq 
\chi_N^{\vec \Lambda}   \le \chi_N^{\Gamma^p(\vec \Lambda)} , ~~~~~
\frac{\chi_{N+1}^{\vec \Lambda} }{\chi_N^{\vec \Lambda} }  \le \frac{\chi_{N+1}^{\Gamma^p(\vec \Lambda)} }{\chi_N^{\Gamma^p(\vec \Lambda)} }  ,
\en
i.e.~the normalisation factor and ratio increase under the application of $\Gamma^p$.

\subsection{Properties of the maximising distribution}
\label{SubSubSecMaximisingDistribution}
The distribution $\vec \Lambda_{\text{max}}(P,\lambda_1)$ that maximises $\chi_N$ for fixed $\lambda_1$ and $P$ is obtained as follows: We maximise the multiplicity of $\lambda_1$ in $\vec \Lambda$, i.e.~$\lambda_1$ is repeated $L-1$ times, with $L=\lceil P/\lambda_1^2 \rceil$. The coefficients then need to fulfil  
\eq 
\lambda_1=\dots = \lambda_{L-1} \ge  \lambda_L \ge \lambda_{L+1} &=& \dots = \lambda_S , \en 
to ensure that  $\vec \Lambda_{\text{max}}(P,\lambda_1)$ be a fixed point of $\Gamma^p$. Again, the distribution coincides with the one found in Ref.~\cite{TichyBosonsBosons}. 


\end{document}